\documentclass[twocolumn,secnumarabic,amssymb, nobibnotes, aps, prd]{revtex4-1}
 
\expandafter\let\csname equation*\endcsname\relax
\expandafter\let\csname endequation*\endcsname\relax
\expandafter\let\csname eqnarray*\endcsname\relax
\expandafter\let\csname endeqnarray*\endcsname\relax
\usepackage{amsmath}
\usepackage{amssymb}
\usepackage{graphicx}
\begin{document}

\title{The effect of polarization entanglement in photon-photon scattering}%

\author{Dennis R\"atzel, Martin Wilkens, Ralf Menzel}
\address{University of Potsdam, Institute for Physics and Astronomy\\
Karl-Liebknecht-Str. 24/25, 14476 Potsdam, Germany}

\begin{abstract}
It is found that the differential cross section of photon-photon scattering is a function of the degree of polarization entanglement of the two-photon state. A reduced, general expression for the differential cross section of photon-photon scattering is derived by applying simple symmetry arguments. An explicit expression is obtained for the example of photon-photon scattering due to virtual electron-positron pairs in Quantum Electrodynamics. It is shown how the effect in this explicit example can be explained as an effect of quantum interference and that it fits with the idea of distance-dependent forces.\\




\end{abstract}

\maketitle

\section{Introduction} 

In the past decades, quantum mechanical entanglement has proven to be a most valuable concept for the understanding of the specifics and the technological applicability of real quantum systems as compared to the physics and the applicability of classical systems. Entanglement is responsible for the violation of the Bell-inequalities in bipartite quantum systems \cite{Bell1964,Aspect1982}. It can also be used for the cramming of two bits of information into one qubit, the speed-up in prime factorization on a quantum computer \cite{Horodecki2009} or applications in quantum optics like quantum imaging, quantum spectroscopy and quantum metrology \cite{Dowling2008}.

In this article we investigate the impact of polarization entanglement in the ``in''-state of two colliding photons on their differential scattering cross section. The paper is organized as follows: In Section \ref{sec:diffcross}, we will derive a general expression for the differential cross section for an initial state that is paramterized so that the degree of entanglement can be varied continuously. In Section \ref{sec:impsymm}, we will use rotational symmetry to reduce the derived expression further. We will give an explicit expression for a specific mechanism in Section \ref{sec:example}; the exchange of virtual electron-positron pairs in Quantum Electrodynamics (QED).   We will interpret the effect of entanglement on the differential cross section from the perspective of quantum interference in Section \ref{sec:intint} and from the perspective of the localization of two-particle states in Section \ref{sec:intforce}.

\section{The differential cross section}
\label{sec:diffcross}

\begin{figure}[h]
\includegraphics[width=6cm,angle=0]{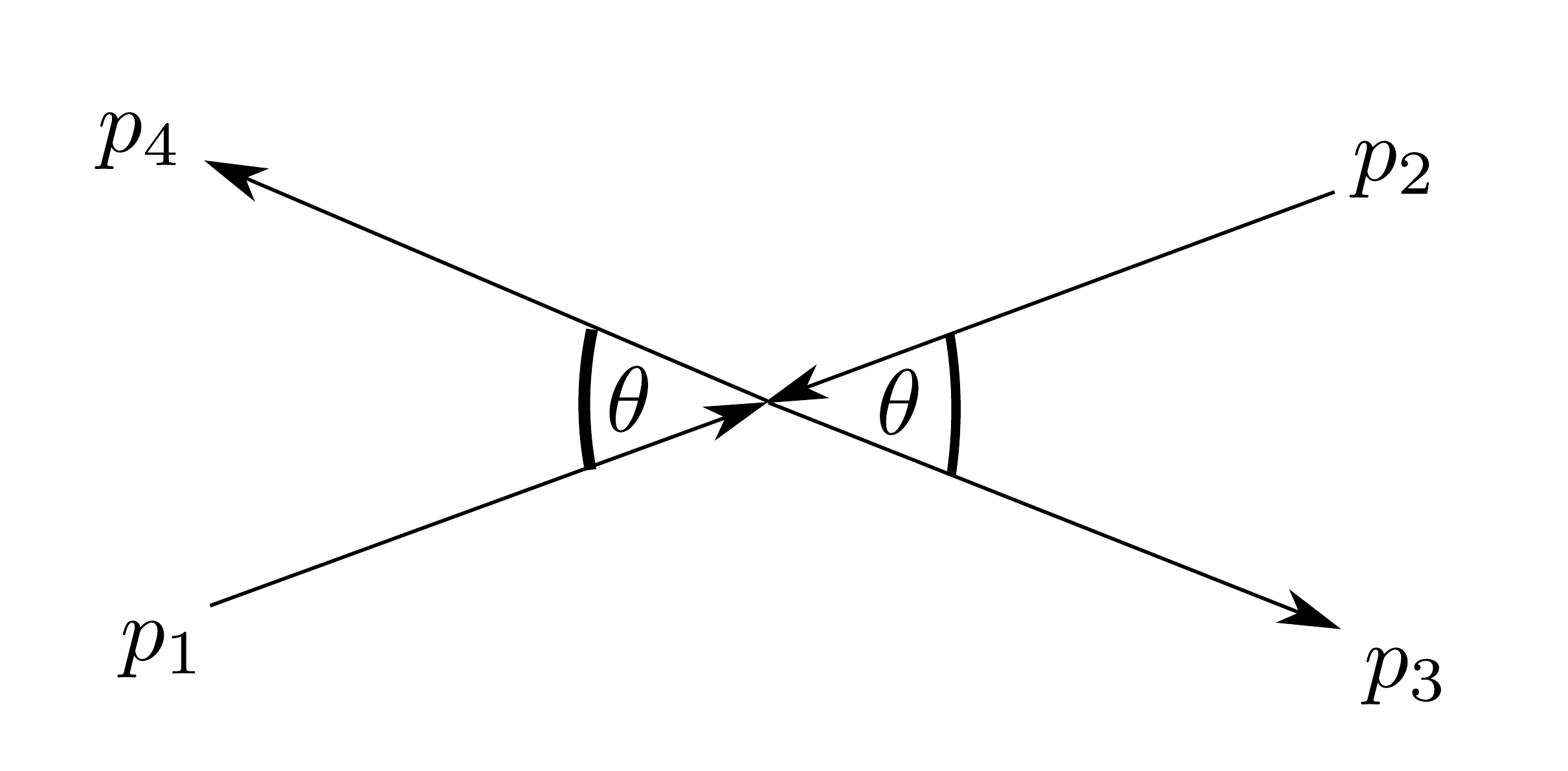}
\caption{\label{fig:int} Photon-photon scattering by the angle $\theta$, in-going momenta $p_1,\,p_2$ and out-going momenta $p_3,\,p_4$. The photons have the polarization $\xi_1,\,\xi_2$ and $\xi_3,\,\xi_4$ respectively, where $\xi_i\in \{1,2\}$.}
\end{figure}

In quantum mechanics, the differential cross section $d\sigma/d\Omega$ gives the probability for an incident particle to be scattered into a specific solid angle element. We assume the background for the scattering of the two photons to be empty Minkowski space. Therefore, we can freely choose the frame of reference for our calculations. A good choice is the center of momentum frame of the two photons where they are counter-propagating. Furthermore, we use the states of linear polarization perpendicular to and along the scattering plane as a basis for the photon polarization. Then, the only remaining variables are the energy of the photons $E$ and the angle $\theta$ between the propagation direction of the incident photons and the propagation direction of the scattered photons (see figure \ref{fig:int}).

We will investigate the effect of entanglement on the differential cross section by considering the following state of two photons of momentum $p_1$ and $p_2$: 
\begin{eqnarray}\label{eq:state}
 	|\Psi\rangle_{\varphi,\rho} &=& \cos\varphi|p_1,1;p_2,2\rangle + e^{i\rho}\sin\varphi |p_1,2;p_2,1\rangle\,,
\end{eqnarray}
where $\xi_1,\xi_2\in\{1,2\}$ in the state $|p_1,\xi_1;p_2,\xi_2\rangle$ label the polarization directions; $\xi_{i}=1$ refers the linear polarization perpendicular to the scattering plane and $\xi_{i}=2$ refers to the linear polarization in the scattering plane.
In equation (\ref{eq:state}), $\varphi\in[0,\pi/2]$ parametrizes the entanglement of the state: for $\varphi= 0$ and $\varphi=\pi/2$, the state is not entangled, and for $\varphi=\pi/4$, it is maximally entangled. The parameter $\rho\in[-\pi/2,3\pi/2)$ governs the relative phase of the superposed states $|p_1,1;p_2,2\rangle$ and $|p_1,2;p_2,1\rangle$. In particular, $|\Psi\rangle_{\pi/4,0}=|\Psi^+\rangle$ and $|\Psi\rangle_{\pi/4,\pi}=|\Psi^-\rangle$ are known as the symmetric and the anti-symmetric Bell state, respectively. Note that we can express the photon momenta via the energy of the photons $E$ as $p_1=p=E/c$ and $p_2=-p=-E/c$, since we are working in the center of momentum frame. 

The differential cross section can be conveniently derived from the scattering matrix $S$ \cite{Itzykson2006}. We define the matrix $M$ such that 
\begin{eqnarray}
	&& \langle p_3,\xi_3;p_4,\xi_4|S|p_1,\xi_1;p_2,\xi_2\rangle\\
	\nonumber && = \mathbb{I}+i\left(2\pi\right)^4\delta^{(4)}(p_1+p_2-p_3-p_4)M_{\xi_1\xi_2\xi_3\xi_4}(E,\theta)
\end{eqnarray}
Due to the rotational symmetry of empty Minkowski space the components $M_{\xi_1\xi_2\xi_3\xi_4}$ are a function of only the energy $E$ and the scattering angle $\theta$ between $p_1$ and $p_3$ (see figure \ref{fig:int}). For the sake of simplicity of the expressions, we will not always write this dependence explicitly in the following.


Considering, for example, the initial state $|\Psi\rangle_{\varphi,\rho}$, we find that the cross section becomes
\begin{eqnarray}
	\nonumber \sigma_{|\Psi\rangle_ {\varphi,\rho}} &=& \frac{c^2\hbar^2}{8E^2}\sum_{\xi_3,\xi_4}d\tilde p_3 d\tilde p_4\times\\ 
	&&\nonumber \times 	 |\cos\varphi M_{12\xi_3\xi_4}  + e^{i\rho}\sin\varphi M_{21\xi_3\xi_4}|^2\times \\
	&&\times (2\pi)^4\delta^{(4)}(p_1+p_2-p_3-p_4)
\end{eqnarray}
where  $d\tilde p_i=d^3p_i/(2p_{i,0}(2\pi)^3)$ is the Lorentz invariant measure for the integration over the spatial components of the four momentum $p_i$. The corresponding differential cross section is given as
\begin{eqnarray}\label{eq:diffcross}
	\frac{d\sigma_{|\Psi\rangle_ {\varphi,\rho}}}{d\Omega} &=& \frac{c^2\hbar^2}{64(2\pi)^2E^2}\times\\
	\nonumber &&\times  \sum_{\xi_3,\xi_4}
	 |\cos\varphi M_{12\xi_3\xi_4} + e^{i\rho}\sin\varphi M_{21\xi_3\xi_4}|^2
\end{eqnarray}
In the following, we will analyze how (\ref{eq:diffcross}) can be reduced due to the rotational symmetry of empty Minkowski space. In section \ref{sec:example}, we will give explicit examples.

\section{The implications of rotational symmetry}
\label{sec:impsymm}

A rotation by 180 degrees in the plane spanned by $p_1$ and $p_3$ rotates $p_2$ into $p_1$ and $p_4$ into $p_3$ while $\theta$ and $E$ stay invariant. Also, the polarization state is invariant under the rotation. It follows that $M_{\xi_1\xi_2\xi_3\xi_4}=M_{\xi_2\xi_1\xi_4\xi_3}$. Because of the conservation of the total angular momentum we have $M_{12\xi_3\xi_4}=0=M_{21\xi_3\xi_4}$ for $\xi_3=\xi_4$. We obtain 
\begin{eqnarray}\label{eq:diffcrossparity}
	\nonumber \frac{d\sigma_{|\Psi\rangle_ {\varphi,\rho}}}{d\Omega} = \frac{c^2\hbar^2}{64(2\pi)^2E^2} \left(|M_{1212}|^2+|M_{1221}|^2+\right.\\
	\left.+2\sin(2\varphi)\cos\rho\, \mathrm{Re}(M_{1212}M_{1221}^*)\right)
\end{eqnarray}
The particles are identical, and we have $M_{\xi_1\xi_2\xi_3\xi_4}(E,\theta)=M_{\xi_1\xi_2\xi_4\xi_3}(E,\pi-\theta)$. Therefore, equation (\ref{eq:diffcrossparity}) can be written only using the matrix element $M_\theta:=M_{1212}$.
\begin{eqnarray}\label{eq:diffcrossparity_2}
	\nonumber \frac{d\sigma_{|\Psi\rangle_\varphi}}{d\Omega} &=&\nonumber  \frac{c^2\hbar^2}{64(2\pi)^2E^2} \left(|M_\theta|^2+|M_{\pi-\theta}|^2+\right.\\
	&& +2\sin(2\varphi)\cos\rho\cos\Delta\beta(\theta)\left.|M_\theta||M_{\pi-\theta}|\right)\,,
\end{eqnarray}
where $\Delta\beta(\theta)$ is the relative phase of the complex functions $M_\theta$ and $M_{\pi-\theta}$, i.e.  $M_\theta M_{\pi-\theta}^*=|M_\theta||M_{\pi-\theta}|e^{i\Delta\beta(\theta)}$. 

Our first observation is that there is no dependence on the degree of entanglement for $\rho=-\pi/2$ and $\rho=\pi/2$, which corresponds to an imaginary factor in front of $|p_1,2;p_2,1\rangle$ in $|\Psi\rangle_{\varphi,\rho}$, while $|p_1,1;p_2,2\rangle$ keeps a real factor. For all other values of $\rho$, we find three regimes of values for $\Delta\beta(\theta)$ in which there is a qualitatively different dependence of the differential cross section on the degree of entanglement of the initial state - expressed through the parameter $\varphi$.  

\begin{enumerate}
\item For $0\le\Delta\beta(\theta)<\pi/2$ and $3\pi/2<\Delta\beta(\theta)<2\pi$, we have $\cos\Delta\beta(\theta)>0$ and the differential cross section (DCS) is larger the stronger the entanglement if $-\pi/2<\rho<\pi/2$ and smaller the stronger the entanglement for $\varphi$ if $\pi/2<\rho<3\pi/2$. In particular, the DCS reaches its maximum for photons in the symmetric Bell state $|\Psi^+\rangle$, and its minimum for photons in the anti-symmetric Bell state $|\Psi^-\rangle$.

\item For $\Delta\beta(\theta)=\pi/2$ and $\Delta\beta(\theta)=-\pi/2$, we have $\cos\Delta\beta(\theta)=0$ and the DCS is independent of the degree of entanglement. 

\item For $\pi/2<\Delta\beta(\theta)<3\pi/2$, we have $\cos\Delta\beta(\theta)<0$ and the DCS is smaller the stronger the entanglement if $-\pi/2<\rho<\pi/2$ and larger the stronger the entanglement if $\pi/2<\rho<3\pi/2$. In this case, the DCS reaches its minimum for photons in the symmetric Bell state $|\Psi^+\rangle$, and its maximum for photons in the anti-symmetric Bell state $|\Psi^-\rangle$.
\end{enumerate}

For right angle scattering, i.e. for the scattering angle $\theta=\pi/2$, we have $\Delta\beta(\theta)=0$, and the DCS becomes
\begin{eqnarray}\label{eq:diffcrossparity_0}
	\frac{d\sigma_{|\Psi\rangle_\varphi}}{d\Omega} &=&  \frac{c^2\hbar^2}{32(2\pi)^2E^2} |M_{\pi/2}|^2\left(1+\sin(2\varphi)\cos\rho\right)\,.
\end{eqnarray}
In case 1., the DCS for right angle scattering vanishes for $|\Psi^-\rangle$ and is larger by a factor two for $|\Psi^+\rangle$ than for not entangled states. In case 3., the DCS for right angle scattering vanishes for $|\Psi^+\rangle$ and is larger by a factor two for $|\Psi^-\rangle$ than for a not entangled states.

In the following, we will give the differential cross section for photon-photon scattering in Quantum Electrodynamics (QED) in the low energy limit as an example.

\section{Example: Quantum Electrodynamics}
\label{sec:example}

\begin{figure}
\includegraphics[width=6cm,angle=0]{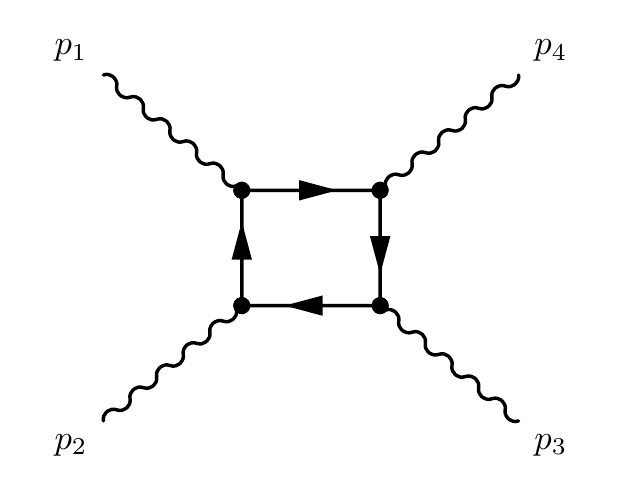}
 	\caption[Feynman diagrams]{One of the Feynman diagrams for the photon-photon scattering due to virtual electrons and positron induced by the photons. The time direction is from the left to the right.}
	\label{fig:elposscatt}
\end{figure}

One possible mechanism for the scattering of two photons is via the vacuum polarization of Quantum Electrodynamics (QED). The process is represented by the diagram in Figure \ref{fig:elposscatt}.
The matrix elements were given first in \cite{Karplus1950} by Karplus and Neumann. The low energy limit in QED photon-photon scattering is applicable for photon energies much below the energy necessary for the creation of real electron-positron pairs. In the low energy limit we have (see also \cite{DeTollis1965,Liang2012})
\begin{eqnarray}\label{eq:Mqed}
	iM_{1212} &=& \frac{4\alpha^2 E^4}{45m^4c^8}\left(31 + 22\cos\theta + 3\cos^2\theta\right)\,,
\end{eqnarray}
where $m$ is the electron mass and $\alpha$ is the fine structure constant. The scattering amplitude in equation (\ref{eq:Mqed}) is always positive and non-zero. Therefore, the matrix elements $iM_{1212}(E,\theta)$ and $iM_{1221}(E,\theta)=iM_{1221}(E,\pi-\theta)$ have the same sign and $\Delta \beta(\theta)=0$ for all scattering angles $\theta$. We obtain case 1. from the previous section and the QED photon-photon scattering has a maximum for the symmetric Bell state $|\Psi^+\rangle$ and a minimum for the anti-symmetric Bell state $|\Psi^-\rangle$. The differential cross section is
\begin{eqnarray}\label{eq:diffcrossqed}
	\frac{d\sigma_{|\psi\rangle_e}}{d\Omega}&=&\frac{\alpha^4}{2\cdot 45^2(2\pi)^2}\frac{\lambda_e^8}{\lambda^6}\times\\
	\nonumber &&\times\left[(1+\sin(2\varphi)\cos\rho)(31+3\cos^2\theta)^2+\right.\\
	\nonumber &&\left.\quad +(1-\sin(2\varphi)\cos\rho)22^2\cos^2\theta\right]\,,
\end{eqnarray}
where $\lambda_e=\hbar/mc$ is the Compton wavelength of the electron and $\lambda=\hbar c/E$ is the wavelength of the two photons. 
\begin{figure}[h]
\includegraphics[width=8cm,angle=0]{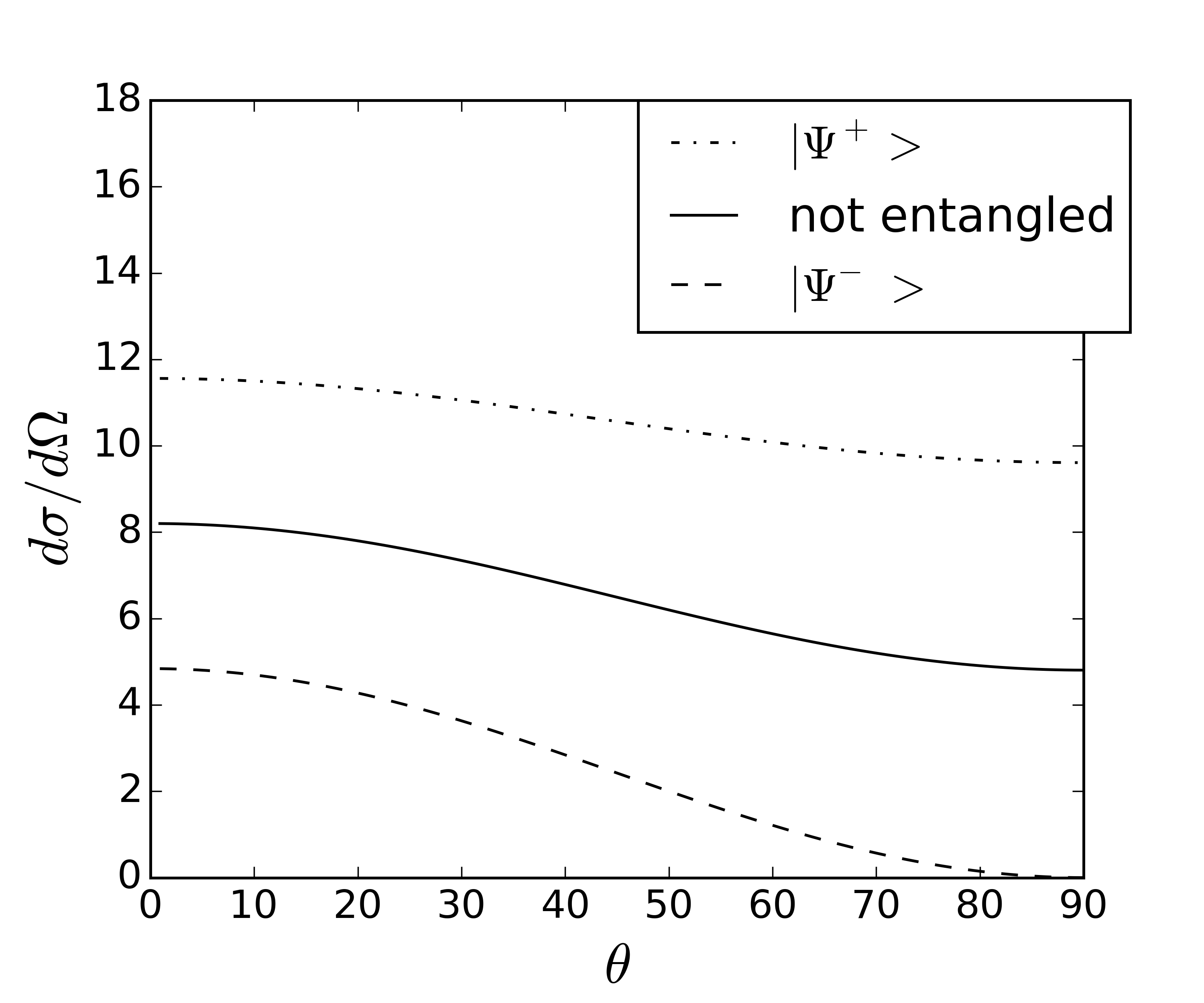}
\caption{\label{fig:plot_diff_qed} The QED differential cross section for not entangled photons and maximally entangled photons in the symmetric Bell state $|\Psi^+\rangle$ and the anti-symmetric Bell state $|\Psi^-\rangle$ in units of $\frac{\alpha^4E^6\hbar^2}{45^2\pi^2m^8c^{14}}\times 10^2$ as a function of the scattering angle $\theta$ from $0^\circ$ to $90^\circ$.}
\end{figure}
The DCS is plotted for the not entangled states and the maximally entangled states $|\Psi^+\rangle$ and $|\Psi^-\rangle$ in Figure \ref{fig:plot_diff_qed}. We see that the effect is already significant for small angles. 

In the following, we will give some numerical examples to discuss the experimental accessibility of the effect in the case of QED photon-photon scattering. For photon wavelengths of about $\lambda=500\mathrm{nm}$, $\lambda=250\mathrm{nm}$ and $\lambda=30\mathrm{pm}$ the polarization averaged cross section for photon-photon scattering of QED is of the order $10^{-72}\mathrm{m}^2$, $10^{-70}\mathrm{m}^2$ and $10^{-47}\mathrm{m}^2$, respectively \cite{DeTollis1965}. Indirect observations of photon-photon interactions were achieved via the detection of Delbr\"uck scattering \cite{Schumacher1975,Akhmadaliev1998} and might be achieved in the near future via the detection of magnetic vacuum birefringence \cite{Bregant2008,Valle2013,Valle2015}. 
The direct QED photon-photon scattering is sometimes called four wave mixing. The most recent experimental upper limits for $\lambda=250\mathrm{nm}$ and $\lambda=30\mathrm{pm}$ are $1.8\times 10^{-52}\mathrm{m}^2$ \cite{Bernard2000} and $1.7\times 10^{-24}\mathrm{m}^2$ \cite{Inada2014}, respectively. However, experiments are planned that might finally see four wave mixing \cite{Lundstroem2006,Tommasini2010,Sarazin2016,Schlenvoigt2016}. Initial states of polarized light are naturally included in these approaches as the light sources are lasers. However, even when polarized photon-photon scattering is observed, the initial states will consist of many photons. It is still a large step to the controlled scattering of entangled two photon states.

\section{Interpretation in terms of interference}
\label{sec:intint}

In this section, we will investigate the reason for the dependence of the differential cross section on the entanglement of the initial state. It is clear from equation (\ref{eq:diffcross}), that the dependence of the differential cross section on the phase factor - which is controlled by the parameter $\rho$ - and on the superposition - which is controlled by the parameter $\varphi$ - is an effect of quantum interference. The amplitude $M_{12\xi_3\xi_4}$ represents the scattering process in which the initial state was $|p_1,1;p_2;2\rangle$ and the final state is $|p_3,\xi_3;p_4;\xi_4\rangle$. The amplitude  $M_{21\xi_3\xi_4}$ represents the scattering process in which the initial state was $|p_1,2;p_2;1\rangle$ and the final state is $|p_3,\xi_3;p_4;\xi_4\rangle$. These different histories interfere to give the scattering amplitude for the state $|\Psi\rangle_{\varphi,\rho}$. 

For QED photon-photon scattering both amplitudes are purely imaginary and their imaginary part has the same sign. Therefore, for $|\Psi^{(+)}\rangle$, they interfere constructively and for $|\Psi^{(-)}\rangle$, they interfere destructively. The DCS for right angle scattering of $|\Psi^{(-)}\rangle$ vanishes because $ M_{12\xi_3\xi_4}$ and $M_{21\xi_3\xi_4}$ must become equivalent for right angle scattering. This is because the history represented by $M_{12\xi_3\xi_4}(E,\pi/2)$ can be transformed into that represented by $M_{21\xi_3\xi_4}(E,\pi/2)$ by a rotation of $180^\circ$ around the axis defined by $p_3$ and $p_4$, and the scattering amplitude is invariant under rotations of the whole setup.

Since all photon states must be symmetric as photons are bosons, the symmetric Bell state $|\Psi^{(+)}\rangle$ is symmetric in the spatial degrees of freedom. This can be seen by formally separating the spatial degrees of freedom off
\begin{eqnarray}
	|\Psi^{(+)}\rangle &=&\frac{1}{\sqrt{2}}(|p_1,1;p_2,2\rangle + |p_1,2;p_2,1\rangle)\\
	\nonumber &=&	\frac{1}{2}(|1\rangle_1|2\rangle_2+|2\rangle_1|1\rangle_2)(|p_1\rangle_1|p_2\rangle_2+|p_2\rangle_1|p_1\rangle_2)\,.
\end{eqnarray}
In contrast, the anti-symmetric Bell state $|\Psi^{(-)}\rangle$ is anti-symmetric in the spatial degrees of freedom:
 \begin{eqnarray}
	|\Psi^{(-)}\rangle &=&\frac{1}{\sqrt{2}}(|p_1,1;p_2,2\rangle - |p_1,2;p_2,1\rangle)\\
	\nonumber &=&	\frac{1}{2}(|1\rangle_1|2\rangle_2-|2\rangle_1|1\rangle_2)(|p_1\rangle_1|p_2\rangle_2-|p_2\rangle_1|p_1\rangle_2)\,.
\end{eqnarray}
\begin{figure}[h]
\includegraphics[width=4cm,angle=0]{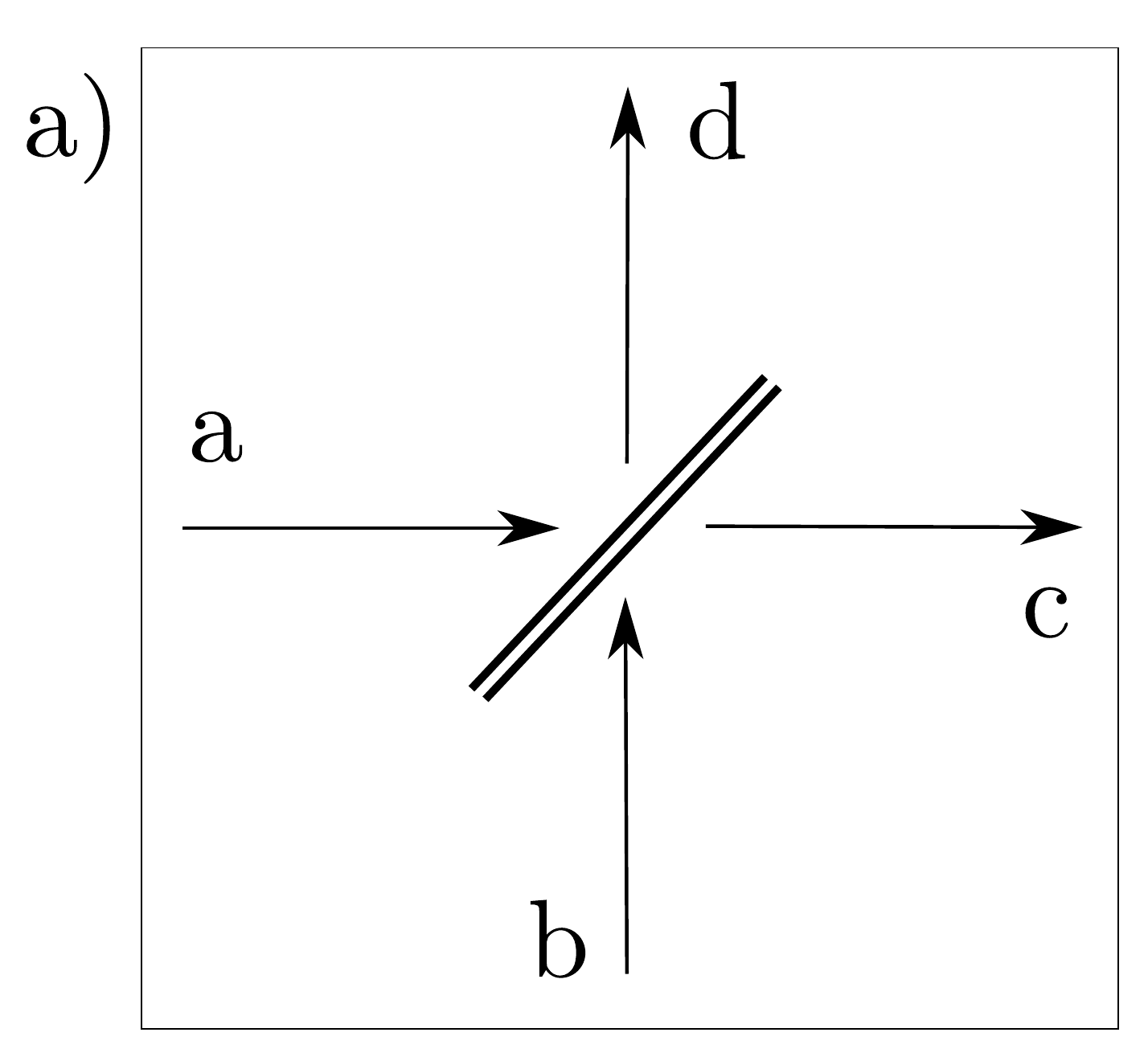}
\includegraphics[width=8.1cm,angle=0]{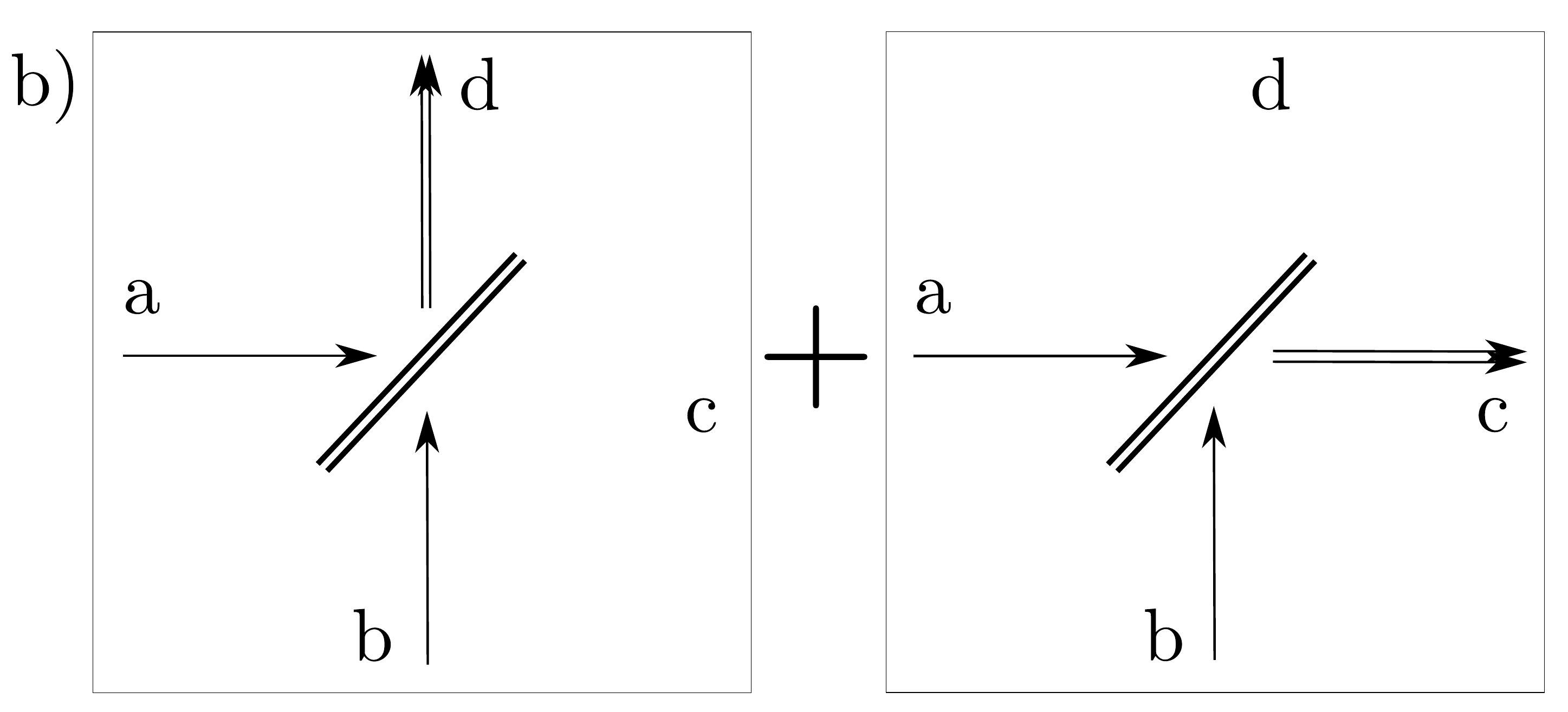}
\caption{\label{fig:beam_splitter} Schematic representation of two photons scattered by a beam splitter. a) In the case of an anti-symmetric Bell state, both photons leave the beam splitter always in different directions. b) In the case of the symmetric Bell state, both photons always leave the beam splitter in a common direction.}
\end{figure}
Therefore, in the case of photon-photon scattering of photons in the symmetric and anti-symmetric Bell state, the interference can be compared to the superposition of two photons at a beam splitter, known from the Hong-Ou-Mandel effect. The situation is illustrated in Figure \ref{fig:beam_splitter}. Let us assume that the two photons are in the anti-symmetric in-state 
\begin{equation}
	|\Psi^{\mathrm{anti-symm}}\rangle=\frac{1}{\sqrt{2}}(|a\rangle_1|b\rangle_2-|b\rangle_1|a\rangle_2)\,,
\end{equation}
where $a$, $b$, $c$ and $d$ label the sides of the beam splitter in counter-clockwise direction. Then, the state after the beam splitter is the anti-symmetric out-state
\begin{equation}
	|\Psi^{\mathrm{anti-symm}}\rangle=\frac{1}{\sqrt{2}}(|c\rangle_1|d\rangle_2-|d\rangle_1|c\rangle_2)\,.
\end{equation}
which means that the photons always appear at different sides of the beam splitter. If the photons are in the symmetric in-state 
\begin{equation}
	|\Psi^{\mathrm{symm}}\rangle=\frac{1}{\sqrt{2}}(|a\rangle_1|b\rangle_2+|b\rangle_1|a\rangle_2)\,.
\end{equation}
the final state is the symmetric out-state
\begin{equation}
	|\Psi^{\mathrm{symm}}\rangle=\frac{i}{\sqrt{2}}(|c\rangle_1|c\rangle_2+|d\rangle_1|d\rangle_2)\,.
\end{equation}
which means that the photons always appear on the same side of the beam splitter.
This is the Hong-Ou-Mandel effect \cite{Hong1987}. It can hardly be interpreted as an interaction of the two photons at the beam splitter since momentum is transferred to the beam splitter and not between the photons when the two photons are co-propagating after the beam splitter. This would correspond to a violation of the conservation of momentum in the two photon system, which can never happen in photon-photon scattering in empty Minkowski space.

\section{Interpretation in terms of distance-dependent forces}
\label{sec:intforce}

In this section, we will use quantum optical measurement theory to interpret the effect of entanglement on the differential cross section in terms of forces between localized particles. The observation that photons in the symmetric Bell state are symmetric in their spatial degrees of freedom and photons in the anti-symmetric Bell state are anti-symmetric in their spatial degrees of freedom already suggests that the different scattering properties of  $|\Psi^{(+)}\rangle$ and  $|\Psi^{(-)}\rangle$ could be interpreted in terms of forces between localized particles.

To elucidate this, we consider delayed coincidence measurements; the detection of one photon at the spacetime point $(t,x)$ and another photon at the spacetime point $(t',x')$.  Let the first photon have the polarization direction $l$ and the second photon the polarization direction $j$. Experimentally, this is easily realized by polarizers in front of the detectors. The probability for the delayed coincidence measurement is then  \cite{Glauber1963} 
\begin{eqnarray}
	P_{i\rightarrow f}\propto \sum_f \left|\langle f| E_l^{(+)}(t,x)E_j^{(+)}(t',x')|i\rangle\right|^2\,,
\end{eqnarray}
where $|f\rangle$ and $|i\rangle$ are the final and the initial state, respectively, and $E_k^{(+)}$ is the positive frequency part of the the $k$-component of the electric field operator
\begin{equation}
	E^{(+)}_k(t,x)=\int d\tilde{p}\, i\omega \,a_{p,k}e^{-\frac{i}{\hbar}p\cdot x + i\omega t}
	 \,.
\end{equation} 
For the initial state $|\Psi\rangle_{\varphi,\rho}$ in (\ref{eq:state}), we find for $l=2$ and $j=1$ 
\begin{eqnarray}\label{eq:coincidence}
	 &&\sum_f \left|\langle f| E_2^{(+)}(t,x)E_1^{(+)}(t',x')|\Psi\rangle_{\varphi,\rho}\right|^2\\
	  \nonumber &=&\langle \Psi|_{\varphi,\rho}\, E_1^{(-)}(t',x')E_2^{(-)}(t,x) E_2^{(+)}(t,x)E_1^{(+)}(t',x')|\Psi\rangle_{\varphi,\rho}\\
	 \nonumber &\propto&\left[1+\sin(2\varphi)\cos\left(\frac{2}{\hbar}p\cdot(x'-x)+\rho\right)\right]\,.
\end{eqnarray}
Note that the time dependence vanishes completely since the mode functions are plane waves.
Since the two photons always have different polarizations, there is only one other non-zero configuration of detected polarizations namely $l=1$ and $j=2$. Going from $l=2$ and $j=1$  to $l=1$ and $j=2$ leads only to an exchange of $x$ and $x'$ in (\ref{eq:coincidence}), since $E_1^{(+)}$ and $E_2^{(+)}$ commute. Therefore, we will only interpret the delayed coincidence measurement probability (\ref{eq:coincidence}) in the following.

The probability in equation (\ref{eq:coincidence}) is modulated by a cosine function that depends on the distance between the two detection points. The size of the modulation depends on the parameter $\varphi$, which controls the entanglement of the state $|\Psi\rangle_{\varphi,\rho}$. In particular, the effect is maximal for entangled states and vanishes for not entangled states. It is more likely to find the two photons at two given points closer than $\pi\lambda/4=\pi\hbar c/4E$ if they are in the symmetric Bell state $|\Psi^{(+)}\rangle$ ($\rho=0$ and $\varphi=\pi/4$) than if they are in any of the not entangled states ($\varphi=0$ and $\varphi=\pi/2$). The parameter $\rho$ - that controls the complex phase between the two not entangled states in equation (\ref{eq:state}) - shifts the cosine function in the direction defined by $p$. The probability of detecting the two photons at two points closer than $\pi\lambda/4$ is decreased for $\rho=\pi$ and $0<\varphi<\pi/2$. The probability to find the two photons at two points in the same plane perpendicular to $p$ vanishes for $\rho=\pi$ and $\varphi=\pi/4$, which is the case of the anti-symmetric Bell state $|\Psi^{(-)}\rangle$.

We conclude that the dependence of the differential cross section on state parameters $\rho$ and $\varphi$ fits with the idea of distance dependent forces. The photons in the symmetric Bell state are much more likely to be found at distances smaller than $\pi\lambda/4$ than not entangled photons which, in turn, are much more likely to be found at distances smaller than $\pi\lambda/4$ than photons in the anti-symmetric Bell state. The QED-interaction depends strongly on the distance between the interacting particle in the low energy limit; the interaction strength decreases exponentially with the distance as the transmitting particles are massive and virtual \cite{Kharzeev2007}. In accordance, photons in the symmetric Bell state deflect each other more than not entangled photons which, again, deflect each other more than photons in the anti-symmetric Bell state.

\section{Conclusions} We derived the differential cross section for photon-photon scattering for a family of initial states with parametrized polarization entanglement which includes the maximally entangled symmetric and anti-symmetric Bell states of linear polarization. We found that there are only three different possible cases for a given scattering angle: 
\begin{enumerate}
\item the scattering is stronger for the symmetric and weaker for the anti-symmetric Bell state than for not entangled photons or
\item the scattering is independent of the degree of entanglement or
\item the scattering is stronger for the anti-symmetric and weaker for the symmetric Bell state than for not entangled photons.
\end{enumerate}
If the scattering amplitude is purely imaginary, we always have case number 1 independent of the scattering angle. 

The result we have derived is limited to the explicit form of our parametrized two-photon state (\ref{eq:state}). To give a general expression for the effect of entanglement in photon-photon scattering, it would be necessary to consider general initial two-photon states. First, we would need a parametrization of all states of a bipartite system like that given in \cite{Tilma2002}. An appropriate measure for the corresponding entanglement could be, for example, the logarithmic negativity \cite{Audenaert2003}. This could be part of a future investigation.

We have given an explicit example of photon-photon scattering mechanisms leading to stronger scattering of entangled photons in the anti-symmetric Bell state: the scattering via virtual electron-positron pairs in Quantum Electrodynamics (QED) in the low energy limit. Another example would be the photon-photon scattering via $W$-bosons in the low energy limit for which the scattering amplitudes can be found in \cite{Jikia1994}. The example of photon-photon scattering in Perturbative Quantum Gravity can be found in \cite{Raetzel2016entgrav}. We considered the experimental feasibility of the detection of the effect in QED, and we found that it is very unlikely to be detected in the near future. The differential cross section for photon-photon scattering  via $W$-bosons is of about the same order as the one for QED, depending on the energy range. The differential cross section for gravitational photon-photon scattering is much smaller than that for QED. 

However, although our results are unlikely to have any experimental implications for photon-photon scattering in vacuum in the near future, their implications in photon-photon scattering in non-linear optical media \cite{Chang2014}, cavities \cite{Hacker2016}, gases \cite{Lukin2000}, plasmas \cite{Platzman1964,Cheng1966,Sodha2008,Mahdy2010} and with Rydberg atoms \cite{Bienias2014,Firstenberg2016} could be observable. 

Independent of its experimental implications, the effect we describe is of general physical interest as it is a general property of photon-photon scattering and entanglement is an important concept for the understanding of the specifics of quantum systems.

In Section \ref{sec:intint}, we investigated the kinematic reason for the effect of entanglement in photon-photon scattering. We argued that the effect can be interpreted as an interference of histories corresponding to the superposed not entangled states. We compared the situation to the situation at a beam splitter, known from the Hong-Ou-Mandel effect.

We used the probability for a two point coincidence measurement, known from quantum optics measurement theory, to show that, in the case number 1, the effect of entanglement on the photon-photon scattering fits well with the idea of localized particles interacting via a force that decays with the spatial distance. To summarize, photons of wavelength $\lambda$ in the symmetric Bell state are much more likely to be found at a mutual distance less than $\pi\lambda/4$ than photons that are not entangled. The latter are again much more likely to be found at a mutual distance less than $\pi\lambda/4$ than photons in the anti-symmetric Bell state. The probability to find photons in the anti-symmetric Bell state at the same point even vanishes. Since the strength of the QED photon-photon interaction depends strongly on the distance of the photons, the corresponding differential cross section for photons in the symmetric Bell state must be increased while that for photons in the anti-symmetric Bell state must be decreased in comparison to the differential cross section of not entangled photons. It can be concluded that this dependence of the differential cross section on the entanglement must be a general property of photon-photon scattering, if the underlying interaction mechanism leads to a sufficiently rapid decrease of the interaction strength with the distance between the photons.

\section*{Acknowledgments}

DR thanks Dan Carney for helpful remarks and discussions and Kiri Mochrie for
proof reading the manuscript and providing writing assistance.


\bibliographystyle{ieeetr} 
\bibliography{entangscatt}

\end{document}